\documentclass[article,floatfix,onecolumn]{revtex4}

\usepackage{graphicx}
\usepackage{epstopdf}
\usepackage{amssymb}
\usepackage{amsmath}
\usepackage{xspace}
\usepackage{color}

\newcommand{\mean}[1]{\langle #1 \rangle}

\begin{document}

\title{The power of a control qubit in weak measurements}
\author{Raul Coto$^{1}$, V\'ictor Montenegro$^{1,3}$, Vitalie Eremeev$^{2}$,
Douglas Mundarain$^{4}$, Miguel Orszag$^{1}$}

\affiliation{$^1$Instituto de F\'{\i}sica, Pontificia Universidad Cat\'{o}lica de Chile,
Casilla 306, Santiago, Chile\\
$^2$Facultad de Ingenier\'ia, Universidad Diego Portales, Av. Ejercito 441,
Santiago, Chile\\
$^3$Department of Physics and Astronomy, University College London, London
WC1E 6BT, UK \\
$^4$Departamento de F\'{\i}sica, Universidad Cat\'{o}lica del Norte, Casilla 1280, Antofagasta, Chile}

\date{September 12, 2016}

\begin{abstract}

In the late 80s, a curious effect suggested by Aharanov,
Albert and Vaidman \cite{Aharonov} opened up new vistas regarding quantum measurements on
weakly coupled systems. There, a combination of a ``weak'' finite
interaction together with a ``strong'' post-selection measurement leads to
an anomalous effect, namely the mean value of a spin-1/2 particle in the
$z-$direction lies outside the conventional spectrum of $\pm$1. Despite
being just a theoretical curiosity, the achieved amplification could be
useful in the realm of sensoring modest quantities below the standard
quantum limit, where they would not be able to be detected otherwise. Hence,
the accurate quantum control of the weak value amplification becomes highly
essential for quantum sensoring and detection.

In this paper, we investigate the quantum control of the weak value
amplification of a qubit system coupled to a meter, via a second
non-interacting qubit, initially quantum correlated with the first one. Our
results show that for weak measurements, the control can be remotely realized via the post-selected state of the second qubit or the degree of
squeezing of the meter. Additionally, in a step towards the study of the
quantum control of the amplification, we can easily manipulate the degree of
quantum correlations between the initial correlated qubits. We find that the
degree of Entanglement has no effect on the quantum control of the
amplification. However, we have found a clear connection between the
amplification and quantum discord like
measurements as well as classical correlations between the qubits. Moreover, we generalize the analysis to two control qubits and we can conclude that the single control qubit scheme is more efficient. Lastly, we suggest an original application of the amplification control protocol on the enhancement of the quantum measurement accuracy, e.g. measuring the relative phase of the post-selected control qubit in a more precise way, as opposed to the no-amplification case.
\end{abstract}

\maketitle

Over the past three decades, important advances have been made using
characteristics of light beams or matter to control the evolution of atomic
and molecular systems. For instance, the development of new and highly coherent laser
sources allow to control molecules in the ground
state \cite{Rabitz}. However, in order to have control over a quantum system is not compulsory to involve external fields, recently were proposed many alternative methods to control \cite{Wu, Kofman, Blok, Torres} and even drive the system to a target state \cite{Wiseman1, Ashhab}.  
Quantum control of physical systems has been a central issue in
recent quantum technology in relation to measurement-based
processes \cite{Wiseman2}, like for example entangling mechanical motion to microwave
radiation \cite{Palomaki}, so for two physical systems, measurement of one system can
determine the state of the other. An interesting control mechanism was recently proposed in an optomechanical system, to control the quantum state of light (single photons) using mechanical variables to monitor a beam splitter \cite {Sahar}, which goes beyond the usual goals of this type of systems, that uses light to control a mechanical resonator.

Quantum Measurement Theory is as old as Quantum Mechanics. The collapse of
quantum states in the measurement process, one of the basic assumptions in
quantum mechanics and put forward by von Neumann in 1932 \cite{von Neumann},
strongly modifies such a state. The question then arises: what would happen
if the interaction responsible for the measurement becomes weaker and
weaker? For weak measurements (WM), a theory was developed by Aharonov and
collaborators \cite{Aharonov}, where the strong impact of the measurement is drastically
reduced. It consists in a gradual accumulation of information during a
finite interaction time between the meter and the system. As a matter of
fact, the state is hardly changed and after such a measurement the system is
left in a state that in general is not an eigenstate of the observable we
are trying to measure, which seems to contradict the basic principles of
Quantum Mechanics. However, this is not so, since the information obtained
after one event is so modest, that many measurement processes are necessary
to actually get information on the system.

In the seminal paper \cite{Aharonov}, Aharonov, Albert and Vaidman (AAV)
showed that the combination of a weak measurement followed by a strong
post-selection measurement may lead to some strange effect, usually referred
to as an anomalous Weak Value Amplification (WVA), anomalous in the sense
that the inferred mean value of the measured system variable lies outside
its range of eigenvalues. The AAV results have been discussed in many papers 
\cite{Duck, Vaidman, Vaidman2, Jozsa, Koike} and also experiments have been
realized and confirmed their predictions \cite{Ritchie, Pryde}. More,
recently, ultra sensitive measurements have been performed \cite{Dixon}, as
well as precision metrology \cite{Zhang} and an exciting experiment on the
observation of the average trajectories of single photons in a two-slit
interferometer \cite{Kocsis}.

In the framework of the AAV approach, the present work proposes to clarify
and resolve three research tasks, which are very important for theory and
experiments in the Quantum Information Science. The main task is devoted to
the effect of control of a quantum system using the \textit{correlations} as resources in the processes of
weak measurements. The second task clarifies which kind of
correlations are indispensable when the WVA occurs. And the third task
deals with the problem of enhancing the amplification effect by squeezing the meter state, making the interaction even weaker. In the following we present our results in detail.

\vspace{10pt}

\begin{figure*}[t]
\centering 
\includegraphics[width=0.8 \textwidth]{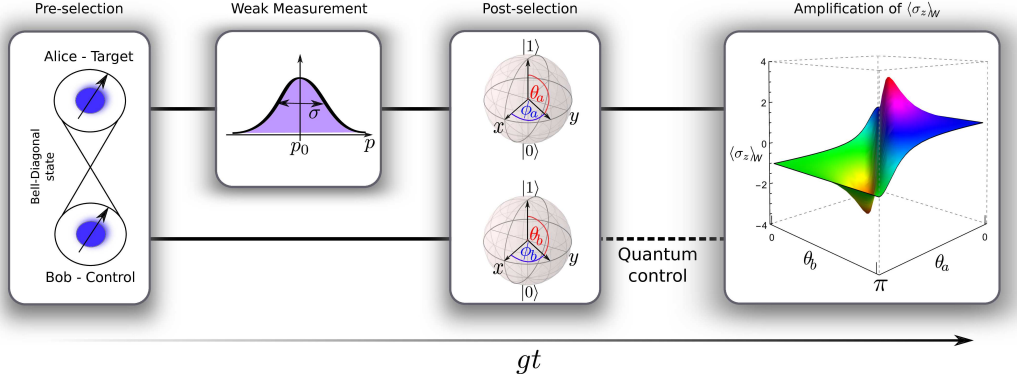}
\caption{Model of weak measurement amplification assisted by quantum
correlated qubits.}
\label{fig1}
\end{figure*}

\textbf{Results}\\
\textbf{Model of Weak Value Amplification assisted by entangled qubits.} Let us
consider two qubits ($a$ and $b$), initially prepared in a Bell Diagonal
(BD) state, $\rho ^{Q}$, such that one of them ($a$) interacts dispersively
with a meter, $\rho^{M}$. The second qubit ($b$) does not
interact at all and is only linked to the system via the quantum
correlations existing between the two qubits, see Fig.(\ref{fig1}). The
Hamiltonian in the interaction picture is 
\begin{equation}
H=\hbar g\sigma _{3}^{a}x,  \label{h1}
\end{equation}%
where $g$ is the coupling strength between the qubit $a$ and the meter; $%
\sigma_{3}$ is the usual spin-$1/2$ Pauli operator in the $z-$direction and $%
x$ denotes the continuous position of the meter. The initial state of the
whole system, i.e., the two qubits together with the meter state is 
\begin{equation}
\rho (0)=\rho ^{Q}\otimes \rho ^{M}=\frac{1}{4}(\mathbb{I}+\sum_{j=1}^{3}c_{j}\sigma
_{j}^{a}\otimes \sigma _{j}^{b})\otimes |\phi \rangle \langle \phi |,
\label{BD}
\end{equation}%
where $\mathbb{I}$ is the identity operator in the two-qubit basis, $\sigma _{j}$ are
the Pauli operators and $|c_{j}| \le1$ are parameters satisfying the positivity of the density matrix. As known, BD states are defined by a set of
three parameters $\{c_1, c_2, c_3\}$ depicted in a three dimensional
tetrahedron, a geometrical representation of the subsets of entangled,
separable and classical states \cite{Luo2, Lang, Liu}. 

Certainly, from the quantum measurement theory, the state of the meter must
be expanded in the opposite conjugate variable appearing in Eq. \ref{h1}, in our
case the momentum subspace $|\phi \rangle =(2\pi \sigma
^{2})^{-1/4}\int_{-\infty }^{\infty }dp\,|p\rangle e^{-\frac{(p-p_{0})^{2}}{%
4\sigma ^{2}}}$, where $\sigma $ and $p_{0}$ are the width and the center of
the Gaussian profile, respectively. Subsequent the time evolution, we
proceed to post-select the target state using a generic qubit state in the
Bloch sphere as $|\psi _{a}\rangle =\cos (\theta _{a}/2)|1\rangle _{a}+\sin
(\theta _{a}/2)e^{i\phi _{a}}|0\rangle _{a}$ (see Fig. \ref{fig1}). Notice
that $|1\rangle $ and $|0\rangle $ are eigenstates of $\sigma _{3}$ with
eigenvalues $1$ and $-1$, respectively. To calculate the post-selected state
of the system $\rho_{\psi_a}=\langle \psi_a |\rho (t)|\psi_a \rangle $, we
make use of the usual translational operator in quantum mechanics, $%
e^{-igtx}|p\rangle =|p-gt\rangle $. Using the above equations and some
algebra one gets

\begin{eqnarray}
\rho _{\psi_a}&=&\frac{1}{4\sigma \sqrt{2\pi }} \int dp\,dp^{\prime }\,e^{-%
\frac{(p-p_0)^2}{4\sigma ^2}-\frac{(p^{\prime }-p_0)^2}{4\sigma ^2}} {\large %
\{ \cos ^2 (\theta _a/2)\rho_{11}^Q}  \vert p-gt \rangle \langle p^{\prime }-gt \vert 
+ \sin ^2 (\theta_a/2) \rho _{00}^Q \vert p+gt \rangle \langle p^{\prime }+gt \vert  \notag \\
&+& \cos (\theta _a/2) \sin (\theta _a/2)[\rho _{10}^Q e^{-i\phi_a} \vert p-gt
\rangle \langle p^{\prime }+gt \vert+h.c.] {\large \} ,}
\end{eqnarray}
with $\rho _{11}^Q= {_a\langle} 1 \vert \rho ^Q\vert 1 \rangle_a =\mathbb{I}^b+c_3\sigma_3^b$, 
$\rho _{00}^Q= {_a\langle} 0 \vert\rho ^Q\vert 0 \rangle_a =\mathbb{I}^b -c_3\sigma_3^b$, and $%
\rho _{10}^Q= {_a\langle} 1 \vert \rho^Q\vert 0 \rangle_a=c_1\sigma_1^b-ic_2\sigma
_2^b$, and $\mathbb{I}^b$ is the identity operator in the $b$-qubit basis.

According to the Eq.(\ref{WV}) in the Sec. Methods, one can easily observe
that by measuring a meter variable one can indirectly evaluate the weak
value of the system variable of interest. Because of this, after the
post-selection, we are interested in the expectation value of the momentum,
which can be found by tracing over the  meter degrees of freedom. To
investigate the effect of the control qubit in the amplification process, we
shall leave the momentum expectation value expression as a function of the
operators acting on the control qubit $b$. Furthermore, we would like to
stress that, since the control qubit $b$ does not interact with target qubit 
$a$ nor with the meter, then the specific time at which one acts on $b$ will
not affect the quantum dynamics.

Next, in order to calculate%
\begin{equation}
\mean p\equiv \frac{\mean{Tr_{M}(\rho _{\psi_a}p)}_b}{\mean{Tr_{M}(\rho_{\psi_a})}_b}
\label{mean_p}
\end{equation}
we derive, after some simple algebra, an expression for $Tr_M(\rho _{\psi_a }p)$, 
yielding the following

\begin{equation}
Tr_M(\rho _{\psi_a }p)=\frac{1}{4}[(\mathbb{I}^{b}+c_{3}\sigma _{3}^{b})\cos ^{2}(\theta
_{a}/2)K_{11}+(\mathbb{I}^{b}-c_{3}\sigma _{3}^{b})\sin ^{2}(\theta
_{a}/2)K_{00}+(c_{1}\sigma _{1}^{b}\cos \phi _{a}+c_{2}\sigma _{2}^{b}\sin
\phi _{a})\sin \theta _{a}K_{10}],
\end{equation}
where the integrals $K_{ij}$, see Methods Eq.(\ref{K10}), are found to be $K_{11}=p_{0}-gt$%
, $K_{00}=p_{0}+gt$, $K_{10}=K_{01}=p_{0}e^{-g^{2}t^{2}/2\sigma ^{2}}$. The expression for $Tr_M(\rho _{\psi_a })$,
denominator in Eq.(\ref{mean_p}), is calculated in a similar way. In fact,
the expression is the same as above, by just replacing $K_{ij}$ by $J_{ij}$,
with $J_{11}=J_{00}=1$ and $J_{10}=J_{01}=e^{-g^{2}t^{2}/2\sigma ^{2}}$.

As  mentioned above, one tries to understand the role of the control qubit $%
b$ in the amplification process. To study this, let us consider two
different approaches. (i) Firstly one traces over the control qubit $b$;
(ii) Secondly one proceeds to perform a projection on the qubit $b$.

In the first case, considering Eq.(\ref{mean_p}), one gets 
$\mean p=p_{0}-gt\cos \theta _{a}$.  
It is easy to see that this WM value does not lead to any amplification
(independent of the initial condition) and the expectation value of the
momentum is bounded by $p_{0}\pm gt$. Furthermore, as known \cite{Mundarain}, coherence plays a significant role in the weak amplification process, thus when tracing over the qubit $b$, one eliminates the coherence in qubit $a$ and therefore the amplification effect is
gone. 

In the second approach, one projects the control qubit $b$
to a similar state as for the qubit $a$, i.e. $|\psi _{b}\rangle =\cos
(\theta _{b}/2)|1\rangle _{b}+\sin (\theta _{b}/2)e^{i\phi _{b}}|0\rangle
_{b}$, so calculating as in \cite{Aharonov} the weak value for the spin
operator, $\langle \sigma _{z}\rangle_{W}\equiv (p_{0}-\mean p)/gt$ by
using the Eq.(\ref{weak value}) in the Sec. Methods. The %
expectation value corresponds to 
\begin{equation}
\mean{\sigma_z}_W = \frac{c_3\cos \theta _b+\cos \theta _a}{1+c_3 \cos \theta
_a\cos \theta _b+e^{-\frac{g^2t^2}{2\sigma ^2}} \sin \theta _a \sin \theta
_{b}(c_1\cos \phi _a\cos \phi _b+c_2\sin \phi _a \sin \phi_b)}.
\label{mean_p_proj}
\end{equation}

This is the principal analytical result of our work for the model of
one control qubit in WM. In the following we analyze some particular cases
such as Bell and Werner states, and thereafter the general BD states. 
\vspace{6pt}

\begin{figure}[t]
\centering 
\includegraphics[width=0.4 \textwidth]{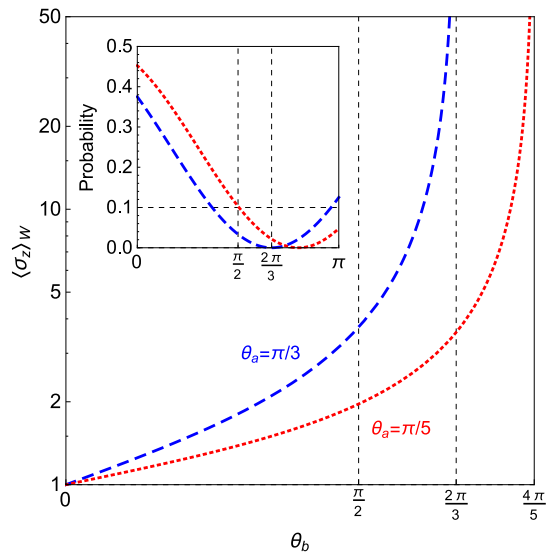}
\caption{The weak value amplification for a \textit{Bell} state, i.e. Eq.(\ref{WVBell}), managed by the projections of the target qubit $a$, and control qubit $b$, with a given probability (Inset). Here $\delta\equiv \phi_{a}+\phi _{b}=\pi$ and $\sigma\rightarrow\infty$.}
\label{fig2}
\end{figure}

\textbf{Weak Value Amplification vs qubit correlations.} 
This section is devoted to the study of the amplification effect via WM
and the quantum correlations shared by the
qubits. The amplification effect in the AAV model appears when the denominator
in the weak value tends to zero, i.e. the pre- and post-selected
states are almost orthogonal. Hence, as a simple and illustrative example
let us consider the case of the two qubits  initially %
prepared in a Bell state, e.g., $|\Phi ^{+}\rangle
=(|0_{a}0_{b}\rangle +|1_{a}1_{b}\rangle )/\sqrt{2}$ ($%
c_{1}=c_{3}=1,c_{2}=-1 $) so Eq.(\ref{mean_p_proj}) is then reduced to 
\begin{equation}
\langle \sigma _{z}\rangle _{W}= \frac{\cos \theta _{b}+\cos \theta _{a}}{%
1+\cos \theta _{a}\cos \theta _{b}+e^{-\frac{g^{2}t^{2}}{2\sigma ^{2}}}\sin
\theta _{a}\sin \theta _{b}\cos \delta }  
\label{WVBell}
\end{equation}
with $\delta =(\phi _{a}+\phi _{b})$. Now, as in \cite{Aharonov} if the
meter state has a large Gaussian spread
distribution on the momentum space, i.e. $\sigma \rightarrow \infty $, one can easily check that there are several different combinations
of the projection angles that allow us to make the
denominator as small as required. For example, we consider the set %
of angles $\{ \delta =\pi, \theta _{a}+\theta _{b}=\pi \}$ which leads to a large amplification with the constraint $\theta _{b}\ne \{ 0, \pi\}$. This constraint comes from the simple fact that for these $\theta _{b}$ values the coherence of qubit $a$ disappears.  The effect of WVA for the Bell state is represented in Fig. \ref{fig2}, as well computing the probability of getting such WVA. The associated probability within the WM limit is calculated as $|\langle \psi _{i}\mid \psi _{post}\rangle |^{2}$, where 
$|\psi _{i}\rangle $ is a Bell state $|\Phi ^{+}\rangle $ and $|\psi _{post}\rangle =|\psi _{a}\rangle \otimes |\psi
_{b}\rangle $, with $|\psi _{a}\rangle $ and $|\psi _{b}\rangle $ being the post selected states for qubits $a$ and $b$. We observe that although exhibiting an infinite amplification, e.g. when $\theta _{b}$ approaches $2\pi/3$ (blue dashed line), which case represents an unphysical state as the probability for this to happen is zero (see inset in Fig. \ref{fig2}). On the other hand, let us look for a realistic/physical scenario, i.e. when a finite amplification with a non-vanishing probability of success is obtained. Fortunately, in the region where an important amplification takes place, the probability is high enough from an experimental point of view. In Fig. \ref{fig2} one finds that a twice amplified expectation value, e.g. red dotted line at $\theta _{b}=\pi/2$, occurs with the probability $\sim10 \%$. 

\begin{figure}[b]
\centering 
\includegraphics[width=0.4 \textwidth]{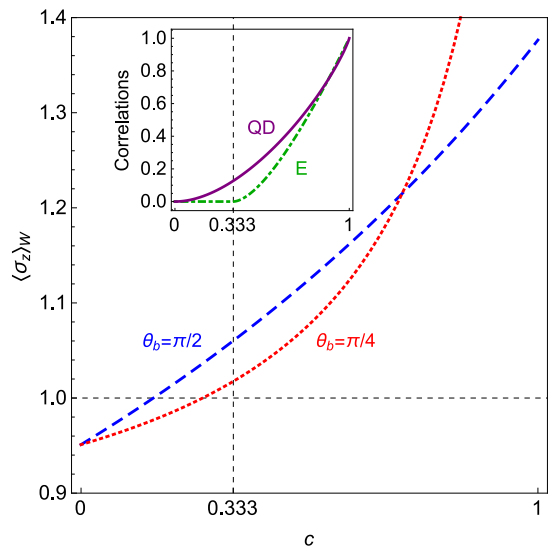}
\caption{The weak value $\langle \sigma _{z}\rangle _{W}$ in Eq.(\ref{mean_p_proj}) computed for a \textit{Werner} state can be controlled by the projection of the control qubit $b$ even for zero Entanglement ($E$) and non-zero Quantum Discord ($QD$)
between the qubits (see Inset). The parameters are $\theta _{a}=\pi/10 $, $\phi %
_{a}=\phi _{b}=0$ and $\sigma\rightarrow\infty$.}
\label{fig3}
\end{figure}
To illustrate more the impact of the control
qubit on the WVA we will proceed to measure the initial
amount of quantum correlations between the qubits. To advance from simple to more  %
elaborated scenarios, firstly we consider a \textit{Werner}
state, i.e. $\rho ^{Q}\equiv \rho _{Werner}$ in Eq.(\ref{BD}). Werner
states are a particular case of BD states when $c_{1}=c_{2}=c_{3}=-c$ %
and they are
defined \cite{Luo2} as $\rho _{Werner}=(1-c)\mathbb{I}/4+c|\Psi ^{-}\rangle \langle
\Psi ^{-}|$, where $|\Psi ^{-}\rangle =(|0_{a}1_{b}\rangle
-|1_{a}0_{b}\rangle )/\sqrt{2}$. 

Furthermore, it is known that Werner states exhibit entanglement if and only if $c\geq 1/3$ 
 (see Fig. 2 in Ref. \cite{Luo2}). Hence, it is clear that the Entanglement of Formation ($E$) vanishes for $%
c<1/3 $, while the Quantum Discord ($QD$) only vanishes at $c=0$.
Following with the
result above, we study the role of quantum correlations in the control for the two-qubit
case. To achieve this, we show in Fig. \ref{fig3} the
amplification of the weak value given in Eq.(\ref{mean_p_proj}) for the Werner state (%
$c_{i}=-c$). There, we have considered two different projections on
the control qubit, $\theta _{b}=\pi/2$ (blue dashed line)
and  $\theta _{b}=\pi /4$ (red dotted line). Without loss of
generality, for both cases we have fixed $\phi _{b}=\phi _{a}=0$, $\theta
_{a}=\pi/10$. For $c < 1/3$, we observe the control of WVA with no entanglement,
therefore the entanglement does not play a relevant role in setting 
up the degree of quantum control. 
Thus, we found in this case that in order to have a control over the target qubit involved in the WVA, one needs to have a resource of quantum correlated states quantified by (in principle) Quantum Discord-like correlation measures rather than non-separability based on, i.e. entanglement. 

For completeness, let us consider the initial uncorrelated state ($QD=0$), $|\varphi \rangle =(|0\rangle_a
+|1\rangle_a )/\sqrt{2}\otimes |0\rangle_b$. For this particular case and following the same procedure as before, it is straightforward to obtain the weak value:
  
\begin{equation}
\mean {\sigma _{z}} _{W}=\frac{\cos \theta _{a}}{1+e^{-g^{2}t^{2}/2\sigma
^{2}}\sin \theta _{a}\cos \phi _{a}}.  
\label{amplif_1qubit}
\end{equation}

One can see from this equation that the amplification of the mean value is %
achievable and it is not influenced by the control qubit %
state, that means that both tracing as well as
projecting the quantum state gives the same result. %
As one would expect, this becomes a clear example of
amplification as in AAV of the first qubit that depends on the ``weakness of the interaction''
with a critical $gt$ value, above which there is no longer amplification,
but there is no control from the second qubit, since they are uncorrelated. In fact, one can find 
 that the amplification tends asymptotically to infinity when $%
\theta _{a}$ approaches $\pi /2$, with $\phi _{a}=\pi $, although the associated probability goes to zero.
To find the ``weak''
interaction within the weak measurement framework, we proceed to
set some routinely values of the quantum dynamics, for instance, $gt=\pi \times 10^{-3}$, and $\sigma =1/2$ (corresponding to a coherent state). These parameters give a
quite accurate approximation of  the case  $%
\sigma \rightarrow \infty $. On the other hand, when we take larger
values of $gt$, the amplification deteriorates. This suggest an optimal
region where the weak interaction takes place. Following a numerical
simulation, one could find that $%
gt\approx 0.3$ corresponds to the threshold where the %
 WVA for the target qubit is achieved for 
 $\sigma =1/2$, $\phi _{a}=\pi $ and $\theta _{a}=1.4$ rad in Eq. (\ref{amplif_1qubit}). 

\begin{figure}[t]
\centering\includegraphics[width=0.4\textwidth]{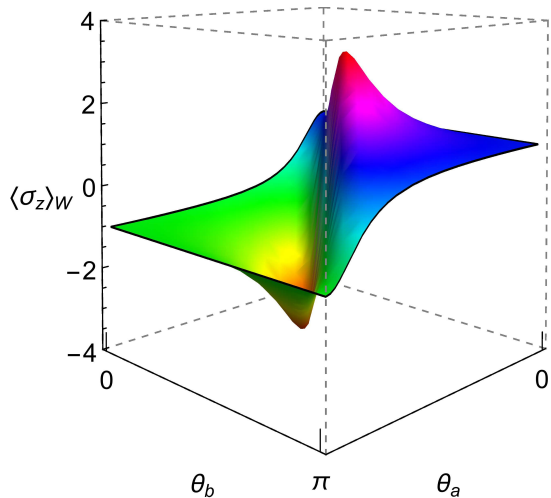}
\caption{Weak value for an initial BD state with $\vec{c}=(-0.95,-0.95,-0.9)$ and
varying the post-selection states for both qubits, e.g. the angles $\protect\theta _{a}$ and $\protect\theta _{b}$. Here $\protect\phi _{a}=\protect\phi _{b}=\protect\pi /4$ and $\sigma\rightarrow\infty$. }
\label{fig4}
\end{figure}
In Fig.(\ref{fig4}) we have depicted the control
of the WVA as a function of the pre- and post-selection
parameters of both qubits. We have
found that, by fixing an initial BD state (i.e. $\vec{c}$) and varying the
angles $\theta _{a}$ and $\theta _{b}$ it is possible to optimize the
amplification effect. Furthermore, it is straightforward to notice that
the behavior of the Entanglement and the $QD$ is similar to the %
previous Werner case (inset of Fig.3). %
Therefore, 
for more general BD states, we have numerically confirmed that %
the Entanglement  plays no role in the
control of the WVA.  
Moreover, we point out on some particular BD states, where the two qubits (target and control) share initially only classical correlations, like states with $\vec{c}=(\pm 1,0,0)$ and $\vec{c}=(0,\pm1,0)$, which lie on the Cartesian axes \cite{Lang}. It is interesting to find that for such states with classical correlations, control over the WVA is possible. We present the details of this issue in the Discussion.

Considering the results of this section, we arrive to the following conclusions:

(i) Essentially, the quantum control consists in 
getting different WVA by manipulating the control qubit through the post-selected angles $%
\theta _{b}$ and $\phi _{b}$. This is the main result %
of our paper, since in the original work of AAV \cite%
{Aharonov} ---where only one qubit is considered--- the amplification
depends only on the strength of the weak measurement, say the
meter spread $\sigma$. In our model, the control qubit $b$ does not
interact with the ``main'' target-meter system and actually it is only connected to %
the qubit $a$ via the initial correlation. This suggests for the first time the idea that the WVA can be remotely switched on and off.

(ii) Further control might be achieved by choosing conveniently the initial BD state, that is the control
via the pre-selection of the two qubits and the quantum or classical correlations between
them.

\vspace{6 pt}

\textbf{Control of amplification with a squeezed meter state.} 
To introduce another degree of quantum control on the WVA, we will proceed to consider a squeezed meter state. Essentially, we are interested in the ratio $gt/\sigma$ again, however this time we will relax the Gaussian coherent meter distribution ($\sigma = 1/2$) with a Gaussian squeezed spread one controlled via $\sigma^{2}=e^{2r}/4$, being $r$ the squeezing parameter. 
\begin{figure}[h]
\centering 
\includegraphics[width=0.4 \textwidth]{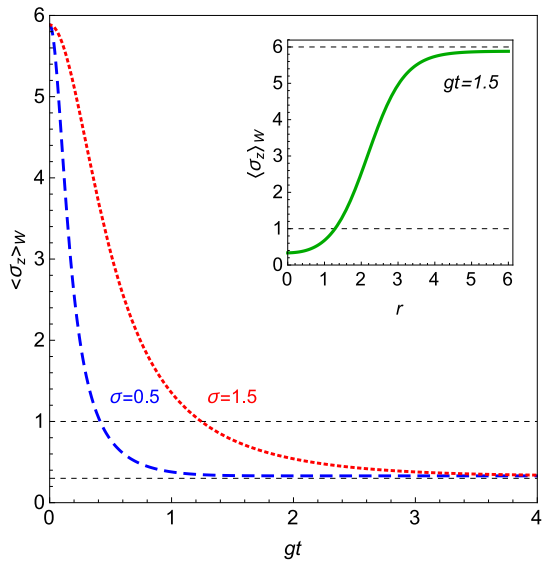}
\caption{The weak value for a \textit{Bell} state, i.e. Eq.(\ref{WVBell}), as function of the dimensionless
time and $\protect\sigma $. The factor $gt$ sets a threshold for having
amplification, which can be moved by tuning $\protect\sigma $, i.e. the
characteristics width of the meter device. In the inset panel, we
consider the case of a squeezed vacuum state, where  $\sigma $ is varied as a function of the squeezed
parameter $r$. Here $\delta =\pi $ and $\theta _{a}=\theta _{b}=1.4$ rad.} 
\label{fig5}
\end{figure}

For a set of values of $\{\theta _{a},\theta _{b},\delta \}$ in the main plot of Fig. \ref{fig5} (blue dashed line), one clearly sees that there is no amplification for the vacuum coherent state ($\sigma = 1/2$) for values higher than ${gt}_{c}\approx 0.4$ ($c$ stands for critical value). On the other hand, considering a squeezed vacuum state for the meter, for instance $\sigma = 1.5$ (red dotted line), we are able to push forward this threshold up, e.g. ${gt}_{c}\approx 1.2$. The horizontal asymptote valued $\langle \sigma _{z}\rangle _{W}\approx0.3$, corresponds to the case where
the interference term (third term in the denominator of Eq.(\ref{WVBell}) 
vanishes and one has no amplification for the chosen angles.

In the inset panel of Fig. \ref{fig5}, we show the variation of the weak value as a function of the
squeezed parameter $r$, for $gt=1.5$. Notice that for $r\lesssim 1.2$ there is no
amplification. However, as we increase $r$ further, the amplification starts to appear saturating its value at $\sim 6$, which is the case when the
exponential in Eq.(\ref{WVBell}) is near to unity.

From Fig. \ref{fig5} one concludes that in the cases where the exponential
term cannot be eliminated, in order to have amplification, the rate $%
gt/\sigma $ should be small (weak measurement constraint). 

\vspace{6pt} 
\textbf{Is multiqubit control more efficient? Three qubits case.} 
For further improvement of our proposal, one may think in adding more qubits
for high quantum control. We will show that in fact this is not
the case, where the case with only one control qubit is the optimal scheme
and that the addition of another one only deteriorates the %
obtained results. Firstly, a direct generalization of the Bell state, $%
|\Phi ^{+}\rangle $, that we used along this paper, is the well-known Greenberger-Horne-Zeilinger ($GHZ$)
state, 
\begin{equation}
|\Psi \rangle _{GHZ}=(|000\rangle +|111\rangle )/\sqrt{2}  \label{GHZ}
\end{equation}%
Then, one proceeds by fixing the parameters
corresponding to the original qubits $a$ (target) and $b$ (first
control qubit), say $\theta _{a}$, $\theta _{b}$ and $\delta$, and
taking values that yield amplification with a finite probability. %
Subsequently, one varies the projection on the third qubit $%
e$ (second control qubit). For a detailed derivation of the results, see
the section Methods. We found numerically, that one can reach higher
values for the amplification, but at the expense of having lower probability as the
previous one qubit control case. Therefore, such an effect does not
lead to any improvement, and even more, it compromises
the experimental success.

Nevertheless, in connection to the three qubit scheme, there is more to
say about the nature of the quantum correlations involved. It have been
proposed and showed that the $GHZ$ state (\ref{GHZ}) has only genuine
three-partite correlations. While a $W$ state, defined as 
\begin{equation}
|\Psi \rangle _{W}=(|100\rangle +|010\rangle +|001\rangle )/\sqrt{3},
\label{W}
\end{equation}
has multipartite correlations, e.g. pairwise Entanglement \cite{Dur}. This
means that for the state (\ref{GHZ}), when tracing out one of the qubits,
the two remaining qubits are not quantum correlated. On the other hand, for (%
\ref{W}), the opposite happens, and the remaining qubits are maximally
quantum correlated. To conclude this section, if one
prepares initially the three qubits as $GHZ$ state, when tracing over one or
two \textit{control} qubits, the amplification is not possible. However, for
the $W$ state, when tracing over only one qubit the amplification
persists. This result proves  that the control of
this type of amplification is intrinsically related to quantum correlations.

\vspace{10pt} 
\textbf{Discussion\\} 
In the present work, we have studied the quantum control of the weak
value amplification (WVA) of a qubit system coupled to a measurement device. On the one hand, a first qubit (target) is directly coupled to the detector device, whereas a second qubit (control) is linked to the former one solely via initial quantum correlations. Motivated by the non-local quantum control of the WVA, we have generalized the single qubit-meter system studied in \cite{Aharonov} towards an entangled multiqubits-meter scheme, being the two-qubit correlated scenario the optimal quantum control case. Particularly, our theoretical analysis shows that correlations of purely quantum nature are of pivotal importance for the WVA, i.e., quantum discord rather than entanglement (correlations based on state separability) is the resource that provides the connection and control over the qubit weak value amplification (Figs. 2-5). For instance, in the case
of the two qubits being in a $Werner$ state, the quantum control prevails
even for the case of zero Entanglement but non-zero Quantum Discord (see
Inset of Fig. 3). However, as mentioned previously, our detailed analysis shows that for some cases where the two qubits are initially classically correlated, the control over the WVA could occur. The explanation of these findings is based on the conclusions presented recently by some of us in \cite{Mundarain}, where it is shown that the presence of coherence in the system is a necessary condition for the existence of WVA, i.e., in our model the measurement of the control qubit $b$ should generate coherence in qubit $a$. As result, for some BD states with only classical correlations we found that the measurement of the qubit $b$ generates the coherence in the qubit $a$, so WVA appears; in the case that the coherence is not generated, the WVA is not reported. On the other hand, for the BD states with quantum correlations (QD $\ne 0$), the coherence is always generated in the system as result of the measurement of the control qubit, hence the protocol of WVA control is robust if the QD is present.

Although quantum discord-like or classical correlations are a necessary condition for the generation of the WVA, we require a projective set of individual local quantum operations on each qubit. For instance, for the pre-selected qubits in a general Bell Diagonal ($BD$) state it is possible to control the WVA %
via qubit projective post-selection measurements (see Fig. 4). %

In the case of achieving WVA, besides the strongly controlled dependence of the amplification due to the phases
($\theta$-azimuthal and $\phi$-polar angles on the Bloch sphere) of the post-selected state of the control and target qubits, we have also found that there is a critical $gt$ value for a fixed Gaussian spread of the meter state $\sigma$, above which no amplification is fulfilled. This remark is in accordance with the original findings shown in Ref. \cite{Aharonov}. There, to gather small amounts of information without perturbing the quantum state, the condition $gt\ll \sigma$ must be attained within the weak measurement framework--- as we also require to approximate the unitary evolution operator up to its first order in $gt/\sigma$. To illustrate this, we have explored different Gaussian spreads of the meter state by varying its
degree of squeezing (see Inset of Fig. 5). We notice that the critical value of $gt$ can be tuned to larger values, as we increase the squeezing parameter $\sigma$.

Lastly, as discussed previously, our amplification scheme relies on several quantum control degrees of freedom, being the projective post-selection measurements the most decisive ones to generate qubit WVA. Of course, one may wonder about the feasibility role of the accuracy in the relative qubit phases, as well as the influence of this in the final amplification. To elucidate this, we would like to draw the attention to one particular but powerful application: enhancement of the control qubit measurement accuracy. In other words, we can rely on the weak value amplification protocol to gain further sensitivity on the post-selected phase $\theta_b$. To accomplish this, we make use of the sensitivity given by:  

\begin{equation}\label{sensitivity}
\eta=\frac{\mean{\sigma_z^{\prime}}_W - \mean{\sigma_z^{0}}_W}{\partial/\partial\theta_b \mean{\sigma_z^\prime}\vert_{\theta_b^0}},
\end{equation}

where $\mean{\sigma_z^{\prime}}_W$, calculated as in Eq. (\ref{WVBell}), is the output value measured for a phase $\theta_b$ that we assume it is slightly displaced from $\theta_b^0=\pi/2$. Needless to say that this small deviation is something that one would expect in any realistic experiment and $\mean{\sigma_z^{0}}_W$ is the theoretical prediction for a perfect measure under ideal conditions. We now proceed to demonstrate that the amplification introduces a higher degree of accuracy. In this direction, we evaluate $\eta$ for two different phases measured on the target qubit, namely $\phi_a=\pi$ (amplification) and $\phi_a=\pi/2$ (no amplification). We fixed the other parameters to be $\phi_b=0$, $\theta_a=\pi/3$ and $\sigma\rightarrow \infty$. Let us say that our meter, for example, can not detect an angle variation of the output below $1\%$. Then, when there is no amplification ($\phi_a=\pi/2$), the sensitivity is about $0.01$. This means that angles below $0.01$ rad can not be resolved in the present configuration. However, when the amplification is switched on ($\phi_a=\pi$) with a non-neglectable probability of $\sim 3.3\%$, the sensitivity corresponds to a value of $0.001$. Thus, we can resolve angles up to $0.001$ rad, being one order of magnitude smaller than the resolution with no amplification.

 We believe that our present work might suggest to develop and/or to implement a new set of experiments and technical tools related
to ultra-small signal amplification via remotely controlled weak measurements by one or more correlated qubits.

\vspace{10pt}

\textbf{Methods} 

\textbf{Brief theory of weak measurements and weak values.} Here we give a
summary of the main results of the AAV's standard approach \cite{Aharonov}.
One begins by preselecting the system $\emph{S}$ in an initial pure state, $%
|\psi \rangle $, such that the state of the system is given as $|\psi
\rangle =\sum_{i}\alpha _{i}|a_{i}\rangle$, where $\{|a_{i}\rangle \}$ is
the set of eigenstates of the system observable $\hat{\mathbf{A}}%
|a_{i}\rangle =a_{i}|a_{i}\rangle$.

On the other hand, let $|\phi \rangle $ denote the wave function of the
measurement apparatus or device detector modeled in terms of continuous
variables $\hat{\mathbf{X}}$ and $\hat{\mathbf{P}}$, such that the initial
detector state may be written as $|\phi \rangle =\int \phi (p)dp\,|p\rangle$%
, with $\phi (p)=(2\pi \sigma ^{2})^{-1/4}e^{-p^{2}/4\sigma ^{2}}$, where $%
\sigma $ being a measure of the quantum fluctuations. In principle, one
could define a WM as the limit when standard deviation $\sigma $ of the
measurement outcome is much larger than the difference between the
eigenvalues of the system. For strong measurements, the opposite is true.

The system-detector Hamiltonian, in the interaction picture, can be written
as 
\begin{equation}
\hat{\mathbf{H}}=g\hat{\mathbf{A}}\otimes \hat{\mathbf{X}},
\end{equation}%
where $g$ is an interaction constant. Thus, the time evolution operator is $%
\hat{\mathbf{U}}(t)=\exp \left\{ -i\frac{gt}{\hbar }\hat{\mathbf{A}}\otimes 
\hat{\mathbf{X}}\right\} $, where $t$ is the interaction time. As a result,
the global system-detector state after interaction is $|\Psi \rangle =\exp
\left\{ -i\frac{gt}{\hbar }\hat{\mathbf{A}}\otimes \hat{\mathbf{X}}\right\}
|\psi \rangle \otimes |\phi \rangle =\sum_{i}\alpha _{i}\int dp\,\phi
(p-gta_{i})\,|p\rangle \otimes |a_{i}\rangle $.

If one takes the WM limit and post-selecting the system state $|\psi
_{post}\rangle $, the measurement device collapses to the state $|\phi
^{\prime }\rangle =\exp (-i\frac{gt}{\hbar }A_{W}\hat{\mathbf{X}})|\phi
\rangle$, where $A_{W}$ is the weak measurement value 
\begin{equation}
A_{W}=\frac{\langle \psi _{post}|\hat{\mathbf{A}}|\psi \rangle }{\langle
\psi _{post}|\psi \rangle }.  \label{WV}
\end{equation}
and the post-selection success probability is 
\begin{equation}
P_{post}=|\langle \psi _{post}|\psi \rangle |^{2}.
\end{equation}
For real $A_{W}$ \cite{Jozsa}, it is easy to show that $\mid A_{W}\mid =%
\frac{\langle \phi ^{\prime }|\hat{\mathbf{P}}\mathbf{|}\phi ^{\prime
}\rangle }{gt}$, a quantity that in many cases has a value outside the range
of the eigenvalues of the observable $\hat{\mathbf{A}}$, in particular in
the limit $\langle \psi _{post}|\psi \rangle \longrightarrow 0.$ If, in
general we write $A_{W}\equiv A+iB$ as a complex number and let $\mathbf{M}$
be any pointer observable, one can easily prove that 
\begin{eqnarray}  \label{eq1}
\langle \mathbf{M} \rangle _{f} &=&\langle \mathbf{M} \rangle _i+ i gtA/
\hbar \langle \hat{\mathbf{X}}\mathbf{M}-\mathbf{M}\hat{\mathbf{X}}
\rangle_i +\frac{gtB}{\hbar }(\langle \hat{\mathbf{X}}\mathbf{M}+\mathbf{M}%
\hat{\mathbf{X}} \rangle _i-2\langle \hat{\mathbf{X}} \rangle _i \langle 
\mathbf{M} \rangle _i),
\end{eqnarray}
with $\langle \mathbf{M} \rangle_i=\langle \phi \vert \hat{\mathbf{M}} \vert
\phi \rangle / \langle \phi |\phi \rangle$, $\langle \mathbf{M} \rangle
_f=\langle \phi ^{\prime } \vert \mathbf{M} \vert \phi ^{\prime } \rangle/
\langle \phi ^{\prime }|\phi ^{\prime }\rangle$, where the $i$ and $f$
indices stand for the initial and final (post selection) states.

In particular, if $A_{W}\equiv iB$ is purely imaginary, then $\langle 
\mathbf{X} \rangle _f=\langle \mathbf{X} \rangle _i+ 2gtB/\hbar Var(\mathbf{X%
})_{i}$. On the other hand, when $A_{W}\equiv A$ is real 
\begin{equation}  \label{weak value}
\langle \mathbf{P} \rangle_f=\langle \mathbf{P} \rangle _{i}-gtA
\end{equation}
\textbf{Solving the integrals.} 

$K_{10} = \frac{1}{\sqrt{2\pi}\sigma} \int_{-\infty}^\infty dp\, (p-gt) e^{-%
\frac{(p-p_0)^2}{4\sigma^2}-\frac{(p-p_0-2gt)^2}{4\sigma^2}}$, by
rearranging the exponential and using the substitution, $\eta =p-p_0$, we
get $K_{10} = \frac{1}{\sqrt{2\pi}\sigma} e^{-\frac{g^2t^2}{2\sigma^2}}
\int_{-\infty}^\infty d\eta \, (\eta-gt+p_0) e^{-\frac{(\eta - gt)^2}{%
2\sigma^2}}$.

Now we introduce a second variable $\xi =\eta -gt$, which leads to the
result 
\begin{equation}
K_{10} = p_0 e^{-\frac{g^2t^2}{2\sigma^2}}  \label{K10}
\end{equation}
The rest of integrals $K_{ij}$ and $J_{ij}$ are calculated similarly. 
\vspace{10pt}

\textbf{Weak measurements with many qubits.}  
Let us see what happens if we include a third qubit $e$ in the model,
i.e. a second control. We are interested in two different types of
tripartite quantum correlated initial states, namely the
GHZ in Eq.(\ref{GHZ}) and W in Eq.(\ref{W}). We firstly focus %
 on the GHZ initial state and we follow the same procedure %
as used to derive the numerator and denominator in Eq.(\ref%
{mean_p}), but written this time as a function of two control qubits, $b$
and $e$ , which gives 
\begin{eqnarray}
tr_M(\rho _{\psi_a}) &=&\frac{1}{4}\{2\Pi _{11}^{be}\cos ^{2}(\theta
_{a}/2)J_{11}+2\Pi _{00}^{be}\sin ^{2}(\theta _{a}/2)J_{00} +[\Pi
_{10}^{be}\sin \theta _{a}e^{i\phi _{a}}J_{10}+h.c.]\},
\end{eqnarray}%
where $\Pi _{ij}^{be}=|ii\rangle \langle jj|$ and $J_{ij}$  %
were defined previously for Eq.(\ref{mean_p}). Either if one
traces over one qubit and project the other, or trace over both $b$ and $e$,
will not get any amplification. 
Nevertheless, projecting on both control qubits we found the
denominator to be 
\begin{eqnarray}
\langle \psi _{be}|tr_M(\rho _{\psi_a})|\psi _{be}\rangle &=&\frac{1}{16}%
\{8\cos ^{2}(\theta _{a}/2)\cos ^{2}(\theta _{b}/2)\cos ^{2}(\theta
_{e}/2)J_{11} +8\sin ^{2}(\theta _{a}/2)\sin ^{2}(\theta _{b}/2)\sin
^{2}(\theta _{e}/2)J_{00}  \notag \\
&+&[\sin \theta _{a}\sin \theta _{b}\sin \theta _{e}e^{i\phi
_{abe}}J_{10}+h.c.]\}
\end{eqnarray}%
where $\phi _{abe}=\phi _{a}+\phi _{b}+\phi _{e}$. One sees that for the
weak regime ($\sigma \rightarrow \infty $) the solution $\{\theta
_{a}=\theta _{b}=\theta _{e}=\pi /2,\phi _{a}=\phi _{b}=\phi _{e}=\pi \}$
leads to amplification (the denominator is zero. However, the strong regime ($\sigma
\rightarrow 0$) will not yield any amplification, as pointed out in \cite%
{Aharonov} for only one qubit.

For the $W$ initial state Eq.(\ref{W}) the denominator reads 
\begin{eqnarray}
tr_M(\rho _{\psi_a}) &=&\frac{1}{6}\{2\Pi _{0000}^{be}\cos ^{2}(\theta
_{a}/2)J_{11} + 2(\Pi _{0101}^{be}+\Pi _{1010}^{be}+\Pi _{0110}^{be}+\Pi
_{1001}^{be})\sin ^{2}(\theta _{a}/2)J_{00}  \notag \\
&+&[(\Pi _{0001}^{be}+\Pi _{0010}^{be})\sin \theta _{a}e^{i\phi
_{a}}J_{10}+h.c.]\},
\end{eqnarray}%
with $\Pi _{ijkl}^{be}=|ij\rangle \langle kl|$. 
Once again, as we found
along this work, when tracing over the two control qubits, the amplification
is annihilated, since the denominator is $1+\sin ^{2}(\theta _{a}/2)\geq 1$ %
. On the other hand, if tracing over $b$ and
projecting on $e$, one gets 
\begin{eqnarray}
&\langle \psi _e \vert tr_{b}[tr_M(\rho _{\psi_a })] \vert \psi _e \rangle =%
\frac{1}{6}\{2\cos ^{2}(\theta _{a}/2)\sin ^{2}(\theta_{e}/2) J_{11}+2\sin
^{2} (\theta _{a}/2)J_{00}+\sin \theta _{a}\sin \theta _{e}\cos
(\phi_{a}-\phi _{e})J_{10}\}
\end{eqnarray}

The denominator does not vanish, but there is still an interference term: $%
\sin (\theta _{a})\sin (\theta _{e})\cos (\phi _{a}-\phi _{e})/2$, which
amplifies the expectation value of momentum, i.e. $ | \langle
p\rangle -p_{0}| /gt \lessapprox 4$. Therefore, for the initial $W$ state
given in Eq.(\ref{W}), one can still find amplification after tracing over
one of the qubits, unlike the GHZ case.

These important differences are related to the quantum correlations, as it
is well known for $GHZ$ state after tracing over one of three qubits, all
the correlations between them are lost, since the \textit{Quantum
Correlations} are purely tripartite. However, for the $W$ state the \textit{%
Quantum Correlations} remain after tracing over one qubit, which is the
reason behind the amplification of the momentum.


\vspace{10pt}

\textbf{Acknowledgments} \newline
M.O. and V.E. acknowledge the financial support of the projects Fondecyt
\#1140994 and Conicyt-PIA anillo ACT-1112 ``Red de analisis estocastico y
aplicaciones". R.C. and V.M. acknowledge the financial support of the
projects Fondecyt Postdoctorado \# 3160154 and \# 3160700, respectively. \vspace{10pt}

\textbf{Author contributions} All the authors contributed equally. 


\begin{thebibliography}{99}

\bibitem{Aharonov} Aharonov, Y., Albert D. Z., \& Vaidman, L. How the result
of a measurement of a component of the spin of a spin-1/2 particle can turn
out to be 100. \textit{Phys. Rev. Lett.} \textbf{60}, 1351-1354 (1988).


\bibitem{Rabitz} Rabitz, H., de Vivie-Riedle, R., Motzkus, M. \& Kompa, K.
Whither the future of controlling quantum phenomena? \textit{Science} 
\textbf{288}, 824-828 (2000).

\bibitem{Wu} Wu, S.,  M\o lmer, K. Weak measurements with a qubit meter. \textit{Phys. Lett. A} \textbf{374}, 34-39 (2009).

\bibitem{Kofman} Kofman, A.G., Ashhab, S. and  \& Nori, F. Nonperturbative theory of weak pre-and post-selected measurements. \textit{Phys. Reports} \textbf{520}, 43-133 (2012).

\bibitem{Blok} Blok,  M. S., Bonato, C., Markham, M. L., Twitchen, D. J., Dobrovitski, V. V. \& Hanson, R. Manipulating a qubit through the backaction of sequential partial measurements and real-time feedback. \textit{Nature Phys.} \textbf{10}, 189-193 (2014).

\bibitem{Torres} Torres, J. P. \& Salazar-Serrano, L. J. Weak value amplification: a view from quantum estimation theory that highlights what it is and what isn't. \textit{Sc. Rep.} \textbf{6}, 19702 (2016).

\bibitem{Wiseman1} Wiseman, H.M. Quantum control: Squinting at quantum systems. \textit{Nature} \textbf{470}, 178-179 (2010).

\bibitem{Ashhab} Ashhab, S. and  \& Nori, F. Control-free control: Manipulating a quantum system using only a limited set of measurements. \textit{Phys. Rev. A } \textbf{82}, 062103 (2010).

\bibitem{Wiseman2} Wiseman, H.M. \& Milburn, G.J. \textit{Quantum Measurement and Control} (Cambridge University Press, 2014).

\bibitem{Palomaki} Palomaki, T.A., Teufel, J.D., Simmonds, R.W. \& Lehnert,
K.W. Entangling mechanical motion with microwave fields. \textit{Science} 
\textbf{342}, 710-713 (2013).

\bibitem{Sahar} Basiri-Esfahani, S., Myers, C.R., Combes, J. \& Milburn,
G.J. Quantum and classical control of single photon states via a mechanical
resonator. \textit{New J. Phys.} \textbf{18}, 063023 (2016).

\bibitem{von Neumann} von Neumann, J. \textit{Mathematical Foundations of
Quantum Mechanics} (Princeton University Press, 1955).

\bibitem{Duck} Duck, I.M., Stevenson, P.M., \& Sudarshan, E.C.G. The sense
in which a ``weak measurement" of a spin-1/2 particle's spin component yields
a value 100. \textit{Phys. Rev. D} \textbf{40}, 2112-2117 (1989).

\bibitem{Vaidman} Vaidman, L. Weak-measurement elements of reality. \textit{%
Foundations of Physics} \textbf{26}, 895-906 (1996).

\bibitem{Vaidman2} Vaidman, L. Weak value and weak measurements. \textit{%
Compendium of Quantum Physics}, Springer(2009), pp.840-842.

\bibitem{Jozsa} Jozsa, R. Complex weak values in quantum measurement. 
\textit{Phys. Rev. A} \textbf{76}, 044103 (2011).

\bibitem{Koike} Koike, T., \& Tanaka, S. Limits on amplification by
Aharonov-Albert-Vaidman weak measurement. \textit{Phys. Rev. A} \textbf{84},
062106 (2007).

\bibitem{Ritchie} Ritchie, N. W. M., Story, J. G. \& Hulet, R. G.
Realization of a measurement of a ``weak value". \textit{%
Phys. Rev. Lett.} \textbf{66}, 1107-1110 (1991).

\bibitem{Pryde} Pryde, G. J., O'Brien, J. L., White, A. G., Ralph, T. C. \&
Wiseman, H. M. Measurement of Quantum Weak Values of Photon Polarization. 
\textit{Phys. Rev. Lett.} \textbf{94}, 220405 (2005).

\bibitem{Dixon} Dixon, P. B., Starling, D. J., Jordan, A. N. \& Howell, J.
C. Ultrasensitive beam deflection measurement via interferometric weak value
amplification. \textit{Phys. Rev. Lett.} \textbf{102}, 173601 (2009).

\bibitem{Zhang} Zhang, L., Datta, A. \& Walmsley, I. A. Precision Metrology
using weak measurements. \textit{Phys. Rev. Lett.} \textbf{114}, 210801
(2015).

\bibitem{Kocsis} Kocsis, S., Braverman, B., Ravets, S., Stevens, M. J.,
Mirin, R. P., Shalm, L. K. \& Steinberg, A. M. Observing the average
trajectories of single photons in a two-slit interferometer. \textit{Science}
\textbf{332}, 1170-1173 (2011).

\bibitem{Luo2} Luo, S. Quantum discord for two-qubit systems. \textit{Phys.
Rev. A} \textbf{77}, 042303 (2008).

\bibitem{Lang} Lang, M.D. \& Caves, C.M. Quantum discord and the geometry of
Bell-diagonal states.\textit{Phys. Rev. Lett.} \textbf{105}, 150501 (2010).

\bibitem{Liu} Liu, B.-H. \textit{et al.}, arXiv:1603.09119 (2016).

\bibitem{Mundarain} Mundarain, D. F. \& Orszag, M. Quantumness of the anomalous weak measurement value. \textit{Phys. Rev. A} \textbf{93}, 032106 (2016).

\bibitem{Dur} D\"{u}r, W., Vidal, G. \& Cirac, J. I. Three qubits can be
entangled in two inequivalent ways. \textit{Phys. Rev. A} \textbf{62},
062314 (2000).
%
%
%
%
%
\end{thebibliography}
\end{document}